**The Mechanism of the Ultra-Fast Crystal Growth of Pure Metals from their Melts**

Gang Sun, Jenny Xu and Peter Harrowell

School of Chemistry, University of Sydney, Sydney NSW 2006 Australia

**A crystal of pure nickel grows from its melt at a rate that reaches 70 m/s. This extraordinary growth rate has led to the suggestion that metallic crystals might provide the next generation of phase change materials. The huge crystal growth rates of metals are a consequence of kinetics without activated control, in sharp contrast to the prediction of the 'classic' theory of crystal growth. While the existence of barrierless growth kinetics is now well established in atomic melts, no physical explanation for the absence of an activation barrier to ordering has been established. It is something of a paradox that diffusion in the liquid metal is governed by thermal activation while the movement of the same atoms as they organize into a crystal is not. In this paper we use computer simulations of crystallization in pure metals to explicitly resolve the origin of the barrierless growth kinetics.**

The kinetics of crystal growth is governed by processes on two quite different length scales. The longer length scale involves the diffusive transport of heat and, in the case of mixtures, species unincorporated into the crystal, from the growing interface [1]. The *intrinsic* growth rate, in contrast, is governed by the microscopic dynamics associated with the cooperative reorganization at the crystal-liquid interface. In computer simulations of crystal growth from the melt, the intrinsic growth rate can be defined as the steady state growth rate when the temperature is constrained to be uniform in space through the application of an appropriate thermostat. Experimentally, this growth rate is most closely realised in the growth rate of a dendrite tip [2].

The steady state intrinsic crystal growth rate ν(T) is described theoretically using a rate model in which the growth rate is expressed as the difference between the rate of crystal addition and subtraction. By invoking microscopic reversibility, this difference can be separated into a thermodynamic term $[1-\exp(\beta\Delta\mu)]$ that accounts for the fraction of crystal additions that are not subsequently reversed by the subtraction process, and a kinetic term, *k(T)*, that represents the rate of addition, i.e.

$$V(T) = k(T)\left[1-\exp(\beta\Delta\mu)\right] \qquad (1)$$

where $\Delta\mu = \mu_{cry}-\mu_{liq}$ is the difference in crystal and liquid chemical potentials and $\beta=1/k_BT$.

By definition, *V* vanishes at $T = T_m$ (a consequence of the equality of rates of crystal addition and subtraction) and the increase in the growth rate on cooling is entirely due to the decrease in the subtraction rate relative to the addition, often referred to as the 'thermodynamic driving force'. In Eq. 1, the kinetics of the ordering process is contained within the magnitude and temperature dependence of the rate of addition *k(T)*. From the literature we can identify three different theories regarding the temperature dependence of *k(T)*: diffusion controlled, collision controlled kinetics and kinetics governed by the relaxation of short wavelength density waves. Diffusion control is described by the Wilson-Frenkel (WF) [3,4] expression $k(T) \propto D(T)/\lambda^2$ where $D$ is the liquid diffusion coefficient and $\lambda$ is an atomic displacement associated with ordering. The WF expression, which predicts a strong decrease in $k$ on cooling due to the activated kinetics of the liquid diffusion, has been found to provide a reasonable description of growth rates of a number of organic molecules [5]. In the collision controlled scenario [6], $k(T) \propto \sqrt{k_B T / M}$ ,where $M$ is the mass of a liquid particle. A simulation study of the crystal growth in an atomic liquid modelled with a Lennard-Jones potential by Boughton, Gilmer and Jackson [7] reported that collision control provided a reasonable fit for the growth of the (110) and (100) crystal surfaces. Finally, we have the time dependent density functional approach, initiated by Mikheev and Chernov [8], and extended using a Ginzburg-Landau (GL) formalism [9,10], where the kinetic coefficient is governed by the relaxation rate of density fluctuations corresponding to reciprocal lattice vector $K_i$ given by $k(T) = \frac{S(K_i)\xi_b}{\tau_L(K_i)N_1A_s}$, where $S(K_i)$ is the liquid structure factor, $N_1$ is the number of principal reciprocal lattice vectors, $\tau_L(K_i)$ is the associated structural relaxation time, the liquid correlation length $\xi_b$ and is the dimensionless anisotropy factor $A_s$as defined in the Methods.

With the development of accurate many body potentials for metals, there have been a number of simulation studies of crystal growth in pure metals [11]. Ashkenazy and Averback [12,13] have reported the temperature dependence of the crystal growth rates for large number of pure metals. They found a turnover in the growth rates at ~ $0.7T_m$ and, largely on the basis of this feature, fitted their growth rates to the WF model. More recently, Mendeleev [14] has also carried out simulations of crystal growth in pure metals and also employed the diffusion controlled theory to model the growth rates at small supercoolings (acknowledging that such a fit may breakdown at higher supercoolings). So the current situation is that, while unchallenged, the collision controlled theory of ref. [6] remains untested and its basis unexplained, while the diffusion controlled theory is being used to model growth despite reservations about its applicability. In this paper, we re-examine and extend these simulations using computational methods developed to study relaxation in supercooled liquids to arrive at an account of the crystal growth rate that is quite different from these previous studies.

We aim to answer the following three questions: i) How does rate of crystal addition of pure metals manage to bypass activation to achieve such enormous rates? ii) What physical parameters determine the rate of crystal addition? iii) What is the physical origin of the turnover in growth rates and subsequent decrease at large supercoolings? To this end we have carried out a series of molecular dynamics simulations of crystal growth for 6 face centered cubic (fcc) forming metals: Al, Ni, Cu, Ag, Pt and Pb. The details of the algorithm and interaction potentials are provided in Methods, along with the details of the constant pressure constraint and local thermostat that ensure the uniformity and isotropy of stress and temperature, respectively, during crystal growth.

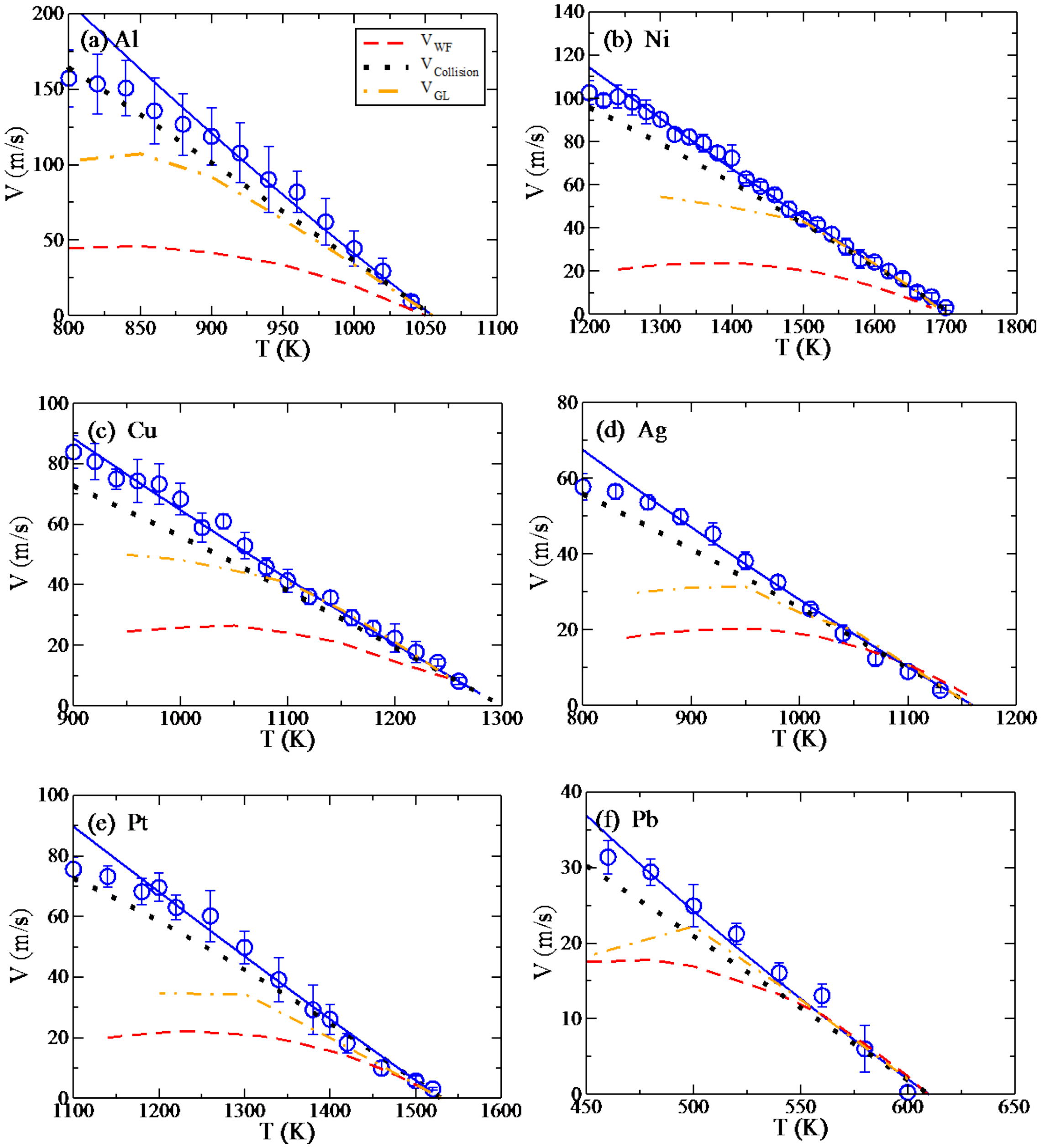

**Figure 1.** Crystal growth rate V as a function of temperature. The average crystal growth rates (blue circles) vs temperature of the surfaces (111) for (a) Al, (b) Ni, (c) Cu, (d) Ag, (e) Pt, and (f) Pb. Included are three theoretical curves, all based on Eq. 1, but with different choices of *k(T)*: the WF prescription $k(T) = l\frac{6D(T)}{\lambda^2}$ (red dashed curve), the collision rate expression $k(T) \propto \sqrt{\frac{3k_B T}{M}}$ (black dots), the expression predicted by Ginzburg-Landau theory $k(T) \propto \frac{S(K_i)}{\tau_L(K_i)}$ (orange dot-dashed curves) and $k(T) = k_0$ (solid blue curve), where $k_0$ is constant, independent of T. Details of the fitting procedure are provided in the Supplementary Information.

| | Al | Ni | Cu | Ag | Pt | Pb |
|---|---|---|---|---|---|---|
| M (g/mol) | 27 | 58.7 | 63.5 | 107.9 | 195.1 | 207 |
| $T_m$ (K) | 1055(±5) | 1700(±5) | 1275(±5) | 1165(±5) | 1530(±5) | 600(±5) |
| $k_0$ (m/s) | 655(±107) | 269(±11) | 212(±17) | 175(±15) | 173(±15) | 142(±10) |
| $\sigma_m$(J/m$^2$) | 11.7 | 27.4 | 16.4 | 13.9 | 15.7 | 4.2 |
| D$_m$($\times10^{-9}$m$^2$/s) | 4.4 | 3.4 | 2.8 | 2.7 | 1.9 | 1.3 |
| $\lambda$ ($\times10^{-10}$m) | 1.17 | 1.47 | 1.36 | 1.41 | 1.38 | 1.2 |
| $l$($\times10^{-10}$m) | 2.3 | 2.03 | 2.09 | 2.36 | 2.26 | 2.86 |

**Table 1.** Properties of the 6 simulated metals in this study. *M* is the atomic mass, $k_0$ is the addition rate obtained by fitting shown in Fig. 1, $\sigma_m$ is the crystal spring constant (see Methods) at $T_m$, $D_m$ is the liquid diffusion coefficient at $T_m$, the displacement length $\lambda$ is explained in the Methods and *l* is the spacing between (111) crystal layers.

In Fig. 1 we plot the crystal growth rates as a function of T for the six pure metals. This data is averaged over 10 individual trajectories. We find that the rate of addition, *k*(T), obtained by fitting Eq. 1 to the simulated growth rates, is essentially independent of T down to supercoolings of ~ 300K for each metal. This means that the addition rate can be well characterised by its value $k_0$, listed in Table 1, evaluated at $T_m$. Over this same range of temperature, the liquid kinetics, as measured by either D or τ, exhibits marked slowing down due to the activated nature of these dynamics (see Supplementary Information). The theories of crystal growth that invoke bulk liquid dynamics, i.e. the WF and GL theories already introduced, reflect this large T dependence. The WF theory deviates from the simulated data even at small supercoolings. In the GL expression, the large T dependence of the time scale $\tau_L$(K) is compensated by the T dependence of 1/S(K), resulting in good agreement between theory and simulated results over small supercoolings. As shown in Fig. 1, this compensation breaks down at larger supercoolings and the GL theory is unable to account for the peak growth rates observed.

The absence of activation in the crystal growth rates is consistent with the experimental observation of Coriell and Turnbull, and their proposal that the ballistic velocity of the particles sets the crystal growth rate is found (see Fig. 1) to provide a reasonable empirical fit to the simulated data. The difficulty of this empirical success is that it lacks any physical basis. In fact, rather than ballistic motion, the short time dynamics of supercooled liquids are well described by the instantaneous harmonic modes [15] associated with fluctuations about the local disordered ground state, the so-called inherent structure first identified by Stillinger and Weber [16]. A possible explanation of ordering without activation is that the liquid adjacent to the crystal surface has a local inherent structure that is already crystalline, but hidden by the thermal particle motion. This possibility would account for the absence of a barrier to ordering since the ordering would, in this case, only require the removal of sufficient heat to make the underlying order apparent. We can explicitly test this idea by carrying out instantaneous potential energy minimizations of the crystal-liquid interface and measuring the movement of the crystal interface arising solely to the minimization. (Details of the minimization procedure can be found in the Method Section.) Any crystallization observed by this procedure represents barrier-less ordering, by construction. To test the method, we have looked at two interfaces of the Lennard-Jones FCC crystal – (111) and (100) – that exhibit very different growth kinetics. As shown by Boughton et al [7], the (100) surface exhibits fast barrierless growth while the (111) surface growth rate is well described by the diffusion controlled WF model. In Fig. 2, we plot the change in interface position due to the minimization of the potential energy of the particles in the interface and liquid. We find that the (111) interface shows only limited ordering on minimization while the (100) interface advances by 5 layers. We conclude that the existence of a crystalline inherent structure for the liquid at the interface correlates with the observation of barrierless crystal growth since the crystal order can be generated without collective reorganization. The absence of an ordered interfacial inherent structure in the case of the (111) surface demonstrates that such ordering is far from inevitable, even in a system as simple as the pure Lennard-Jones liquid.

Having established that the observation of a crystalline interfacial inherent structure does correlate with the observation of barrier less growth, we can now apply the methodology to the metal crystal-liquid interfaces. We find a significant advance of all the metal (111) interfaces when minimized. As an example, we plot in Fig. 3a the advance of the Cu (111) interface by 3 layers due to energy minimization alone. The analogous behaviour of the other 5 metals are presented in the Supplementary Information. We argue that the demonstration that the liquid adjacent to the interface have crystalline ground states explains how crystal growth in these metals can proceed without activated kinetics even as diffusion in the liquid does involve barrier crossing.

The demonstration that we can observe crystal growth, in the case of the pure metals, via the purely deterministic process of energy minimization offers a valuable window onto the microscopic mechanism of ordering. The displacements associated with ordering during energy minimization represent a minimal configuration path from liquid to crystal, shorn of the ‘unnecessary’ thermal fluctuations. The picture of ordering revealed by this process renders explicit the atomic ‘shuffling’ that Jackson [17] proposed as the mechanism of fast crystal growth. As shown in Fig. 3b, the crystal-melt interface advances during minimization via highly collective movement of the particles in the liquid-like regions confined between aligned crystal domains.

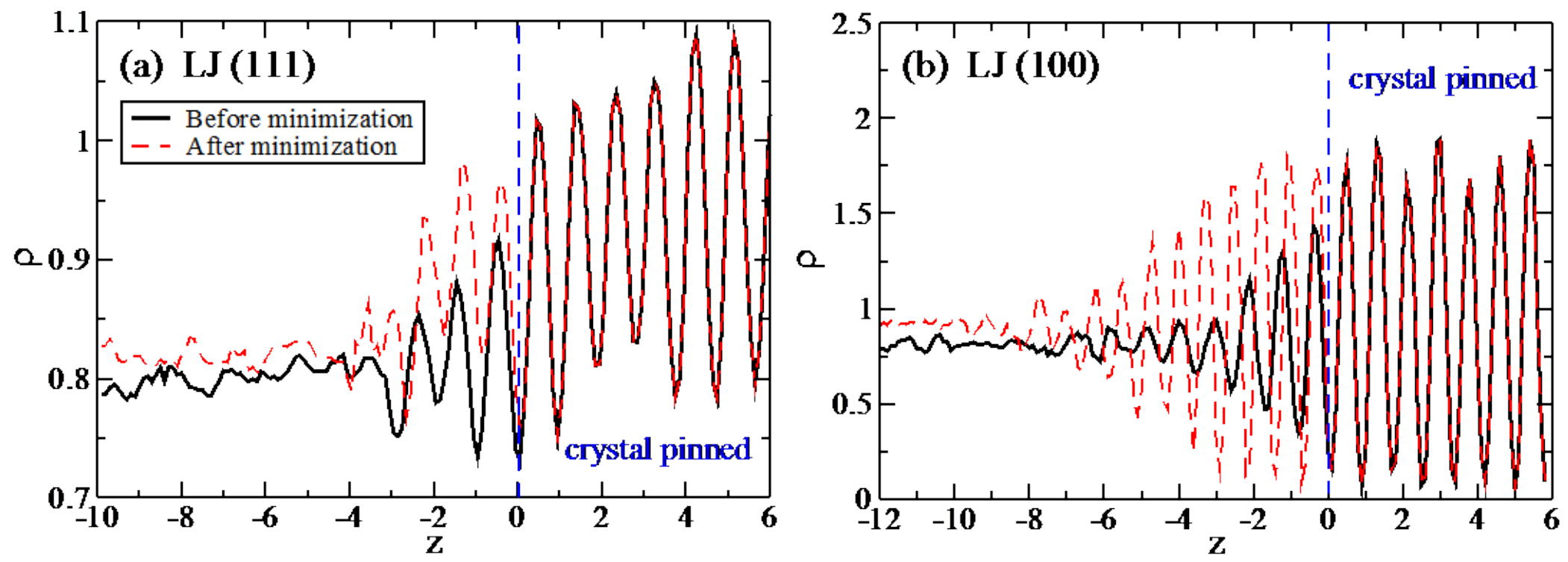

**Figure 2.** The advancement of the Lennard-Jones crystal-liquid interface during energy minimization. The Lennard-Jones models crystal-liquid interface as measured by particle number density, ρ, before and after potential energy minimization at the melting temperature $T_m$=0.58 for (a) (111) and (b) (100) interfaces. The (111) interface forms little additional order during minimization, consistent with the importance of diffusion in its growth kinetics, while the (100) interface exhibits a significant advance (~ 5 layers), behaviour consistent with the differences in growth kinetics, barrier controlled and barrierless, respectively. See Methods section for more details. The blue dashed lines in (a) and (b) mark the position of crystal pinned during energy minimization.

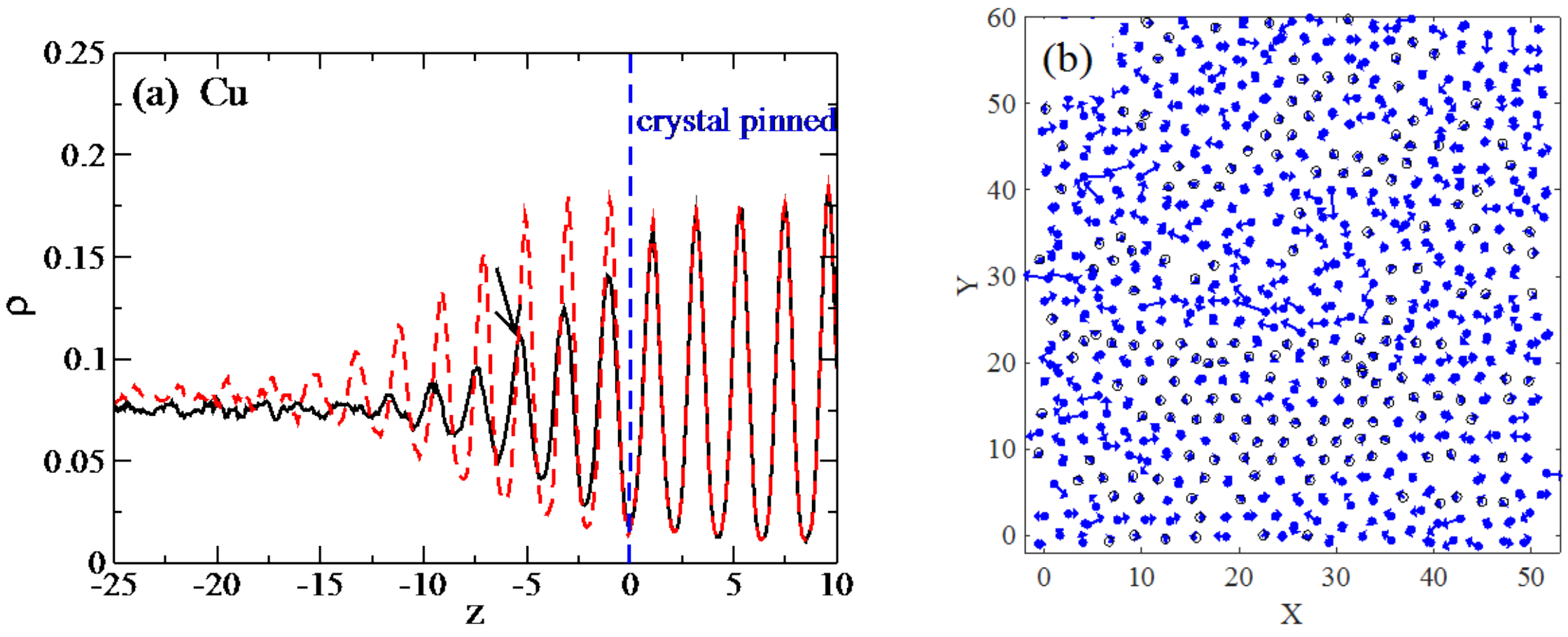

**Figure 3.** The ordering process of the Cu crystal-liquid interface during energy minimization. The Cu crystal-liquid interface as measured by particle number density, ρ, before and after potential energy minimization at T=1260K~ $T_m$ for (a). The advance of the interface corresponds to about 3 crystal layers. The blue dashed line marks the position of crystal pinned during energy minimization. (b) a representation of particle positions in a layer of the Cu interface (indicated by an arrow in (a)) and the displacements associated with ordering during energy minimization. The particles in crystal environments are indicated by open circles and those in disordered environments by blue filled circles.

| | Al | Ni | Cu | Ag | Pt |
|---|---|---|---|---|---|
| $k_0^{theory}$ | 655 | 269 | 212 | 175 | 173 |
| $k_0^{theory} \times r$ (WF) | 149 | 289 | 322 | 343 | 219 |
| $k_0^{theory} \times r$ (GL) | 705 | 735 | 99 | 602 | 936 |
| $k_0^{theory} \times r$ (Collision) | 522 | 399 | 373 | 277 | 234 |
| $k_0^{theory} \times r$ (Vibration) | 520 | 489 | 383 | 304 | 230 |

**Table 2.** The comparison of the simulated kinetic coefficient $k_0^{sim}$ with the theoretical estimates $k_o^{theory}$ from the WF, GL, thermal collision and vibrational addition models. In each case $k_0^{theory}$ has been scaled by a multiplicative factor $r = \frac{k_o^{sim}(Pb)}{k_o^{theory}(Pb)}$ based on the values for *Pb* to simplify the comparison of the predicted trends across the metals.

The preceding analysis provides compelling evidence that the liquid adjacent to the growing interface shares a common local potential minimum with the crystal. This suggests that the characteristic rate with which interfacial configurations (such as those depicted in Fig. 3b) are explored should correspond to the characteristic frequency of the interfacial instantaneous phonons of the interface. We propose to approximate this quantity by using the analogous vibrational frequency of the crystal. This proposal, the reader should note, is in clear contrast to the standard assumption that it is a *liquid* rate that sets the timing for growth. Here we introduce the Einstein frequency of the crystal $\sqrt{\sigma_m / M}$ (where the force constant $\sigma_m$ at T = $T_m$ is defined in the Methods and its value provided in Table 1) and propose that the vibrational addition rate is given by $k_0 = \frac{l}{2\pi}\sqrt{\frac{\sigma_m}{M}}$. In Fig. 1 we plot the crystal growth rate as predicted by a constant rate of addition $k(T) = k_0$, as proposed in our vibrational model. We find that the vibrational theory manages to capture a similar behavior to that expressed in the collisional theory, but without requiring the assumption of ballistic motion. In Table 2, we present four theoretical predictions of $k_0$: the vibrational addition rate, the collisional addition rate, WF kinetic coefficient and that due to the GL theory, against the simulated values. To make the comparison meaningful, we have scaled each theory value by $r = \frac{k_o^{sim}(Pb)}{k_o^{theory}(Pb)}$ such that the scaled theory value is set equal to the simulated rate coefficient for the case of Pb (i.e. the smallest value $k_0$). We find that all theories exhibit significant scatter about the simulation data. The vibrational addition and thermal collision expressions do, however, correctly reproduce the simulated ranking of the values of $k_0$ for the different metals, unlike the WF and GL theories.

At supercoolings greater than ~ 300K, we observe an apparent turnover and eventual decrease in the growth rate of all the metals with decreasing T. The behavior for Cu, which is shown in Fig. 4a, is representative to what is seen in the other 5 metals (provided in the Supplementary Information). This behavior is similar to that reported previously [12,13] where the turnover was used to support the idea of diffusion control and the use of the WF model. Here we shall demonstrate that this apparent turnover is a consequence of the kinetic instability of the liquid with respect to crystallization and, hence, the low temperature data should not be interpreted as a steady state crystal growth rate.

Quenches of the homogeneous liquid reveal that crystallization takes place at a reproducible temperature (see Fig. 4b) that we shall refer to as $T_{sp}$, a temperature which marks the effective end of metastability of the supercooled liquid. (Formally, metastability ends with the disappearance of the liquid minimum in the free energy. In practice, the supercooled liquid ceases to be well defined once the rate of crystal nucleation exceeds the rate of liquid relaxation. This latter condition we refer to as the 'kinetic instability'.) We find that $T_{sp}$ is relatively insensitive to cooling rate above some threshold value of this rate (see Fig. 4b insert). Values for $T_{sp}$ for each of the metals in this study are presented Table 3 along with $T_{max}$, the temperature of the maximum in the growth rate. Comparison of the values $T_{max}$ and $T_{sp}$ reveal that the two temperatures are essentially identical for each metal studied. We conclude the turnover of the ultra-fast crystal growth of the pure metals is a consequence of the onset of unstable growth of local crystal order throughout the liquid.

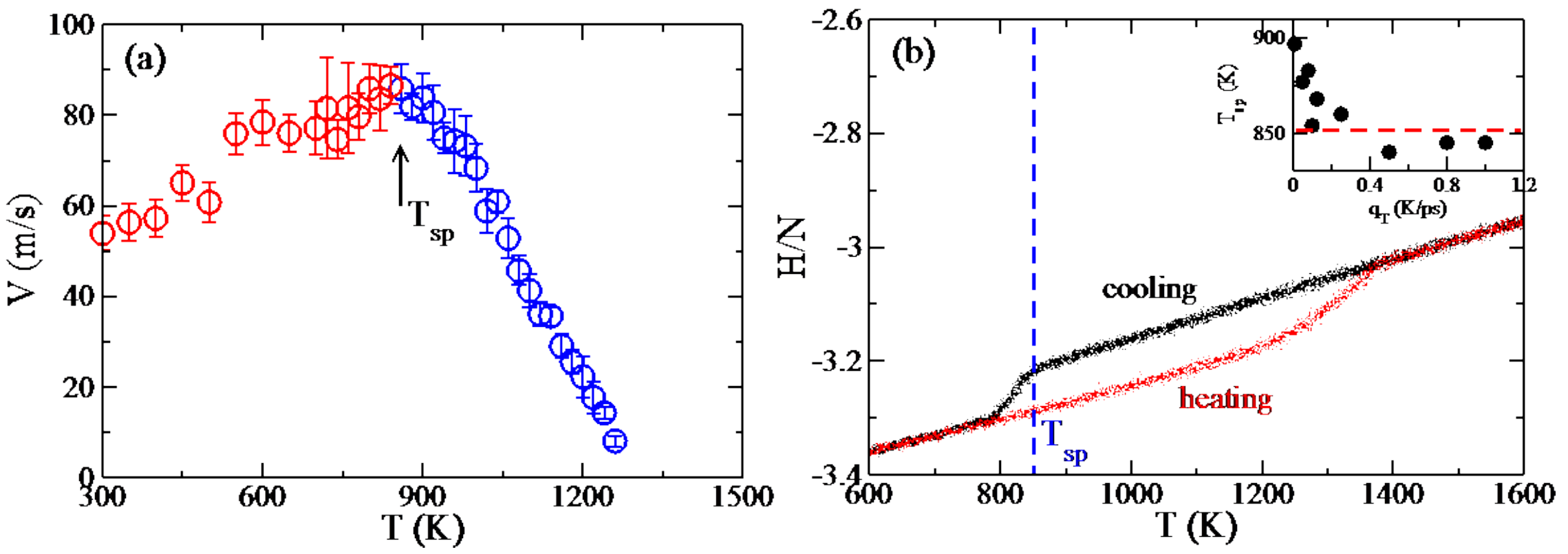

**Figure 4.** The relation between the turnover temperature of crystallization velocities and the kinetic instability temperature. (a) Plot of the average crystal growth rate (blue) vs temperature of the Cu (111) surface and the apparent growth rate (red) below the kinetic instability temperature $T_{sp}$ (see text), and (b) plot of enthalpy H vs T for quenches of Cu identifying the kinetic spinodal temperature $T_{sp}$. Insert: the value of $T_{sp}$ as a function of quench rate $q_T$ demonstrating that the above some threshold cooling rate, $T_{sp}$ is roughly independent of $q_T$. The reported value of $T_{sp}$ is that associated with the plateau indicated by the horizontal dashed line.

The origin of the change in ordering mechanism at $T_{max}$ can be confirmed by inspecting the evolution of a selected layer, located initially in the liquid, well in advance of the interface, during crystallization. As shown in Fig. 5, for $T > T_{max}$, the crystal order parameter $\overline{Q}_6$ (see Methods) remains small and constant until it abruptly rises to the crystalline value, marking the passage of the crystal front. Below $T_{max}$, we find $\overline{Q}_6$ increases continuously from the start of the run with no sign of metastability. In Fig.5 and Fig.S9 (see Supplementary Information), we can see that spontaneous ordering occurs in the liquid well in front of the crystal/liquid interface. The apparent growth rate below $T_{sp}$ is the time average of the intermittent movement of crystal front in the presence of the freezing liquid. This apparent rate is time dependent (due to the freezing of the liquid in advance of the interface) and of no particular physical significance for the overall crystallization kinetics. Previously, Ashkenazy and Averback [13] had dismissed liquid instability as being responsible for the turnover in the growth rate. Their argument was based on equating the absence of a heat signature with an absence of ordering. As is discussed in detail in the Supplementary Information, the relation between order, as measured by $\overline{Q}_6$ , and enthalpy is highly non-linear; a substantial increase in order incurs little enthalpy loss until the final ordering where almost of the enthalpy of fusion is released. This means that a supercooled liquid can order sufficiently to interfere with the propagation of a crystal interface without releasing much heat.

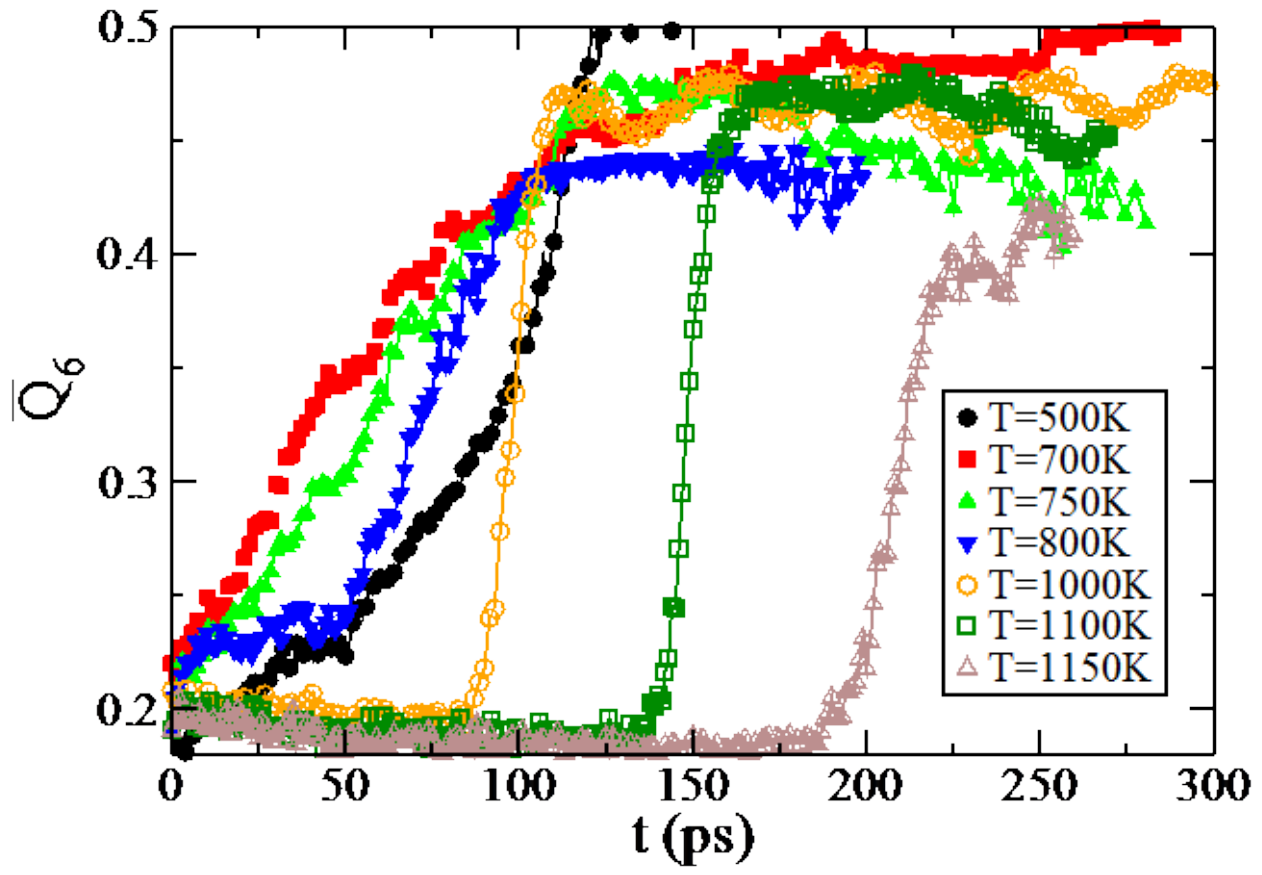

**Figure 5.** Time evolution of a selected layers of liquid Cu during crystallization at a range of temperatures. Note the disappearance of metastability as T < 852K, the value of $T_{sp}$ for Cu.

| | Al | Ni | Cu | Ag | Pt | Pb |
|---|---|---|---|---|---|---|
| $T_m$ (K) | 1055(±5) | 1700(±5) | 1275(±5) | 1165(±5) | 1530(±5) | 600(±5) |
| $T_{max}$ (K) | 760 | 1140 | 840 | 750 | 1100 | 400 |
| $T_{sp}$ (K) | 742 | 1165 | 852 | 787 | 1073 | 400 |

**Table 3.** Properties of the 6 simulated metals in this study where the temperatures $T_m$, $T_{max}$ and $T_{sp}$ are explained in the text.

In conclusion, we have demonstrated that the ultra-fast crystal growth in pure metals is governed by a barrierless ordering process that is made possible by the capacity of the crystal interface to impose a crystalline inherent structure (i.e. the local potential energy minimum) onto the adjacent liquid. The large interface widths that characterise the crystal interfaces in the pure metals appear to be an important factor in establishing the ordered interface inherent structures. Through the introduction of the inherent structure of the crystal-liquid interface, we have presented a general algorithmic approach to assessing the degree to which an interface influences the local ground state of the adjacent liquid in simulations and, hence, the kinetics of its motion. Having established the crystalline character of the ground state of the interfacial liquid, we have introduced a expression for the rate coefficient, the vibrational addition expression, based on the insight that ordering at the interface is dominated by a crystalline inherent structure. This crystal-based time scale provides a better account of the T dependence of $k_0$ than those based on liquid dynamics while avoiding the unphysical neglect of particle interactions that underlies the thermal collision model. Further improvement in the theory of ultra-fast crystal growth will proceed, we suggest, through consideration of anharmonic perturbations of the crystal dynamics in the interfacial region.  Finally, we find that the apparent turnover the growth rate at large supercoolings previously reported is not a consequence of intrinsic liquid dynamics but, rather, a result of the kinetic instability of the liquid with respect to crystallization. Below $T_{max}$, ordering is no longer determined by the propagation of a single crystal front and so this quantity loses any particular physical significance.

Having established a clear physical account of what the upper bound of possible crystal growth rates as represented by the pure metals, the opportunity now exists to explain the slower growth rates observed from more complex liquids in terms of the explicit manner in which the complexities (e.g. composition fluctuations or molecular orientation) disrupts the order of the interface inherent structures and so necessitates activated reorganization prior to ordering.

## Methods

To investigate the crystal growth rate of (111) faces of pure metals, we first created the crystal/liquid interface samples. The initial simulation box contains 10x10 transformed unit cells in xy plane and 40 transformed unit cells along z direction, with total 64000 atoms. Then, the crystal particles of 10 unit cells in the middle were pinned, and the rest crystals were melt into liquids at high temperatures (above melting temperatures). Periodic boundary conditions were applied in three spatial directions, and the NPT ensembles were performed to study the growth rate. Finally, we set the pinned particles free and relaxed the whole crystal/liquid systems for 300ps (with time step 1 fs ) at the melting temperature and at pressure =0, controlled independently in three directions. During crystal growing, to release the latent heat as fast as and keep the temperature uniform in space, the whole crystal/liquid systems were divided into narrow slabs with width Δd =1nm, and the temperatures were separately controlled. For all the simulations, temperatures were controlled via Berendsen thermostat and pressures by Nose-Hoover barostat. To detect the crystal front, the crystal/liquid interface was identified via the spatial distribution of order parameter $\overline{Q_{6,i}}$ [18,19], which was the average form of local bond order parameter $Q_{6,i}$ [19] over all its neighbors and itself. We measured the crystal growth rate by detecting the width of the crystal. The particle with $\overline{Q_{6,i}}$ > 0.95*$\overline{Q_{6,i}}\big|_{crystal,\,T=T_m}$ is considered as crystal. The energy minimization of crystal-liquid interface was performed using the conjugate algorithm [20]. During energy minimization, the crystal particles were fixed, and the free boundary condition was applied along the crystal growth direction (z direction). The length λ is obtained by calculating the displacement of the liquid particles during crystallization via energy minimization. Specific, λ is the average displacement distance of the particles in the layer with the maximum change $\overline{Q_{6,i}}$ during energy minimization for the configurations at the melting temperatures. For crystal lattice vibration, the spring constant at melting temperature, $\sigma_m$, is obtained from the plateau of mean squared displacement, $<\Delta r^2>$, using the fluctuation relation, $\sigma_m = \frac{3k_B T_m}{<\Delta r^2>}$. In the kinetic coefficient predicted by the Ginzburg-Landau theory [10,11], the liquid relaxation time $\tau_L(K_i)$ is defined as the time when the self intermediate scattering function $F_s(k,t)=1/e$. The liquid correlation length $\xi_b$ corresponds to the inverse half-width of the liquid structure, $\xi_b = \left[-S''(K_i)/2S(K_i)\right]^{1/2}$. The dimensionless anisotropy factor is defined as $A_s = \frac{1}{N_1}\sum_{k_i} \xi_b / \xi_{\bar{k}_i}$, where $\xi_{\bar{k}_i}$ is the effective widths of density wave profile. Under the isotropic approximation, the corresponding anisotropy factor is $A_s = \frac{\sqrt{3}}{6} u_s^2$, where $u_s$ is the amplitude of density wave in solid. Embedded-atom method (EAM) potentials were used to describe the interaction between atoms for all the metals, Al [21], Ni [22], Cu [22], Ag [22], Pt [22], and Pb [23]. For the Lennard-Jones model, the interaction potential is given by $u(r) = 4\varepsilon\left[(\frac{\sigma}{r})^{12} - (\frac{\sigma}{r})^6\right]$, and truncated at r=2.5σ. The units are scaled by parameters σ, ε and m. The melting temperature for FCC crystal in Lennard-Jones model is ≅0.58 [24]. All the simulations were performed via the code package LAMMPS [25].

## Data and code availability

The data sets generated and/or analysed and the codes that the evidence is generated from during the current study are available from the corresponding author upon reasonable request.

## Acknowledgements

We gratefully acknowledge programing assistance from Chunguang Tang. This work has been supported by a Discovery grant from the Australian Research Council.

## Author contributions

G.S. carried out the majority of calculations, designed and prepared all of the figures, contributed to the writing of the paper, and was involved in assessing the outcomes of the various computational approaches. J.X. carried out a number of preliminary calculations. P.H. conceived the project, designed the overall computational strategy, assessed the outcomes and wrote the paper.

## Additional information

Supplementary information is available in the online version of the paper. Reprints and permissions information is available online at www.nature.com/reprints. Correspondence and requests for materials should be addressed to P.H.

## Competing financial interests

The authors declare no competing financial interests.

## Supplementary Information

**The Mechanism of the Ultra-Fast Crystal Growth of Pure Metals from their Melts** by Gang Sun, Jenny Xu and Peter Harrowell

**Contents**

♣

## 1. The Fitting of Theoretical Curves in Fig. 1

The WF fits in this work and that in Ashkenazy and Averback [1,2] and Mendelev [3], are different. In Refs. [1-3], the version of the WF expression used is

$$V(T) = C\exp(-Q/k_BT)\left[1-\exp(-\Delta\mu(T)/k_BT)\right] \tag{S1}$$

where $\Delta\mu(T)$ is the difference of chemical potentials in liquid and crystal, and Q is the diffusion activation energy. $\Delta\mu=\mu_l-\mu_c$, is obtained by assumption $\Delta\mu(T)=(T_m-T)\ \Delta h_m/T_m$, where $\Delta h_m$ is the enthalpy of fusion per particles at melting temperature $T_m$. Eq. S1 essentially assumes that the liquid diffusion is Arrhenius and, more importantly, the activation energy Q is treated as a fitting parameter rather than being determined independently from direct evaluations of the diffusion constant D. The satisfactory fits (for small supercoolings) of Eq. S1 in these references are the result of treating both C and Q as fitting parameters. In this paper, we have retained the Wilson-Frenkel original description, crystal growth velocity is given by,

$$V(T) = l\frac{6D(T)}{\lambda^2}\left[1-\exp(-\Delta\mu(T)/k_BT)\right] \tag{S2}$$

where D(T) is the diffusion constant, λ is the displacement during crystallization and *l* is the lattice spacing normal to the surface. In Fig. 1, the WF curves we display are the result of using Eq. S2 and the explicitly calculated values of the diffusion constant of the pure metal liquids, D(T), and the values of λ obtained explicitly by calculating the displacement of the liquid particles during crystallization via energy minimization. We believe that the WF curves shown in Fig. 1 represent an accurate representation of the growth rates predicted by this theory.

Based on the collision-limit theory, the crystal growth velocity is given by

$$V(T) = C\sqrt{\frac{3k_BT}{M}}\left[1-\exp(-\Delta\mu(T)/k_BT)\right] \qquad \text{(S3)}$$

where M is the mass of liquid atom. In Fig. 1, we determined the parameter C by fitting Eq. S3 to the crystallization velocities at temperatures close to the melting point.

According to the Ginzburg-Landau theory, the expression of crystal growth velocity is

$$V(T) = \frac{S(K_i)\xi_b}{N_1\tau_L(K_i)A_s}\left[1-\exp(-\Delta\mu(T)/k_BT)\right] \qquad \text{(S4)}$$

where $S(K_i)$ is the liquid structure factor, the relaxation time $\tau_L(K_i)$, the liquid correlation length $\xi_b$ and the dimensionless anisotropy factor $A_s$ are defined in the Methods.

## 2. Additional Data on the Simulated Metals

**Table S1** The fractional density difference between liquid and crystal at melting temperatures, and the width $d_{eq}$ of the crystal/liquid interface equilibrated at the melting point, for all six metals. Besides, the heat of fusion corresponding to the potential and the magnitude of λ at the melting temperature is also presented below.

| | Al | Ni | Cu | Ag | Pt | Pb |
|---|---|---|---|---|---|---|
| $\Delta\rho/\rho_{liquid}$ | 4.8% | 6.96% | 4.25% | 7.38% | 2.49% | 3.32% |
| $d_{eq}$ (Å) | 12 | 8 | 10 | 12 | 10.5 | 14 |
| $\Delta h_m/k_BT_m$ | 1.06 | 1.3 | 1.21 | 1.1 | 1.44 | 0.93 |
| $\lambda / l$ | 0.5 | 0.7 | 0.65 | 0.6 | 0.6 | 0.42 |

**Table S2** The diffusion constant D at selected temperatures for all six metals are presented here. The spacing between (111) crystal layers *l* and the displacement length $\lambda$ are presented in the **Table 1**.

| | | | | | | |
|---|---|---|---|---|---|---|
| **Al: T (K)** | 800 | 900 | 1000 | 1050 | 1150 | 1200 |
| **D** ($\times 10^{-9}$ $m^2/s$) | 1.61 | 2.13 | 3.9 | 4.35 | 5.87 | 6.4 |
| **Ni: T (K)** | 1240 | 1300 | 1400 | 1500 | 1600 | 1680 |
| **D** ($\times 10^{-9}$ $m^2/s$) | 0.96 | 1.23 | 1.73 | 2.25 | 2.9 | 3.37 |
| **Cu :T (K)** | 950 | 1000 | 1050 | 1100 | 1150 | 1200 |
| **D** ($\times 10^{-9}$ $m^2/s$) | 1.00 | 1.25 | 1.56 | 1.8 | 2.1 | 2.25 |
| **Ag: T (K)** | 840 | 900 | 1000 | 1060 | 1100 | 1160 |
| **D** ($\times 10^{-9}$ $m^2/s$) | 0.713 | 0.988 | 1.55 | 1.94 | 2.24 | 2.69 |
| **Pt: T (K)** | 1140 | 1200 | 1300 | 1400 | 1460 | 1520 |
| **D** ($\times 10^{-9}$ $m^2/s$) | 0.63 | 0.8 | 1.11 | 1.47 | 1.7 | 1.93 |
| **Pb: T (K)** | 440 | 500 | 560 | 600 | 620 | 640 |
| **D** ($\times 10^{-9}$ $m^2/s$) | 0.49 | 0.77 | 1.11 | 1.32 | 1.49 | 1.57 |

### 3. The Relation between Enthalpy and Structure

Previously, Ashkenazy and Averback [2] had dismissed liquid instability as being responsible for the turnover in the growth rate. They argued that since the growth rate as measured by the propagation of the order front and the growth rate as determined by the rate of heat release coincided, it followed that an instability of the liquid phase could not be responsible for the observed growth behaviour. The weak point of this argument is that it equates the absence of a heat signature with the absence of ordering. As shown in Fig. S1 the relation between order, as measured by $\overline{Q_6}$, and enthalpy is highly non-linear; a substantial increase in order incurs little enthalpy loss until the final ordering where almost of the enthalpy of fusion is released. This means that a supercooled liquid can order sufficiently to interfere with the propagation of a crystal interface without releasing much heat.

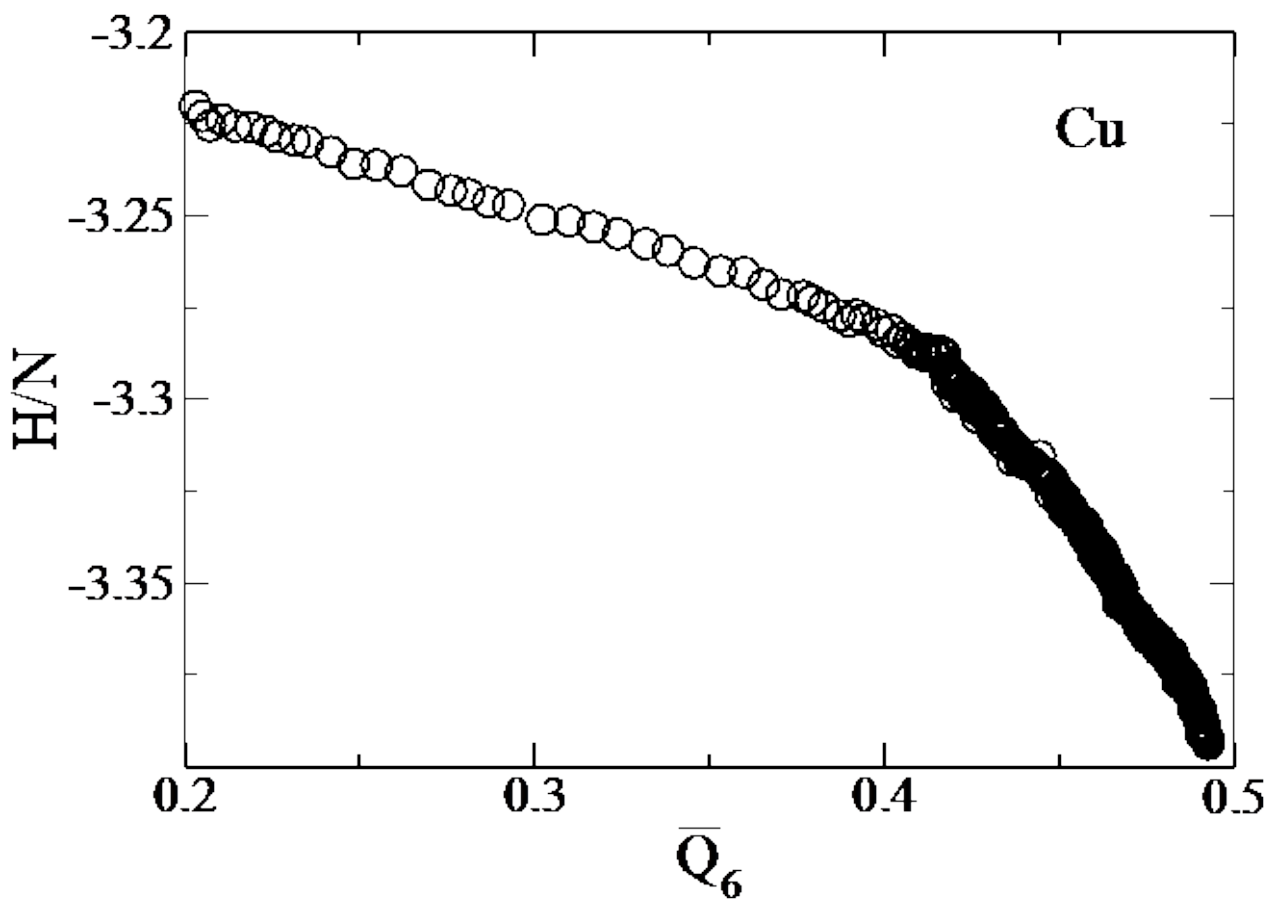

**Figure S1.** The relation between enthalpy and order in Cu during spontaneous crystallization near $T_{sp}$. Note that the principle release of heat only occurs as $\overline{Q}_6$ approaches the crystal value.

## 4. Additional Data on Crystal Growth due to Minimization

In Fig. 2, the effect of minimization on the (111) and (100) interfaces for the Lennard-Jones system was shown using the density profiles. We can also represent the difference between original and minimized surfaces using the order parameter $\overline{Q}_6$ as shown in Fig. S2. While both representations show the same behavior, we chose to use the density profiles because we felt it provided a more direct insight into the degree of crystalline ordering. The order parameter $\overline{Q}_6$, while excellent for distinguishing crystal from liquid at a common temperature, is not so useful when comparing a high temperature structure and one that has been quenched to T = 0.

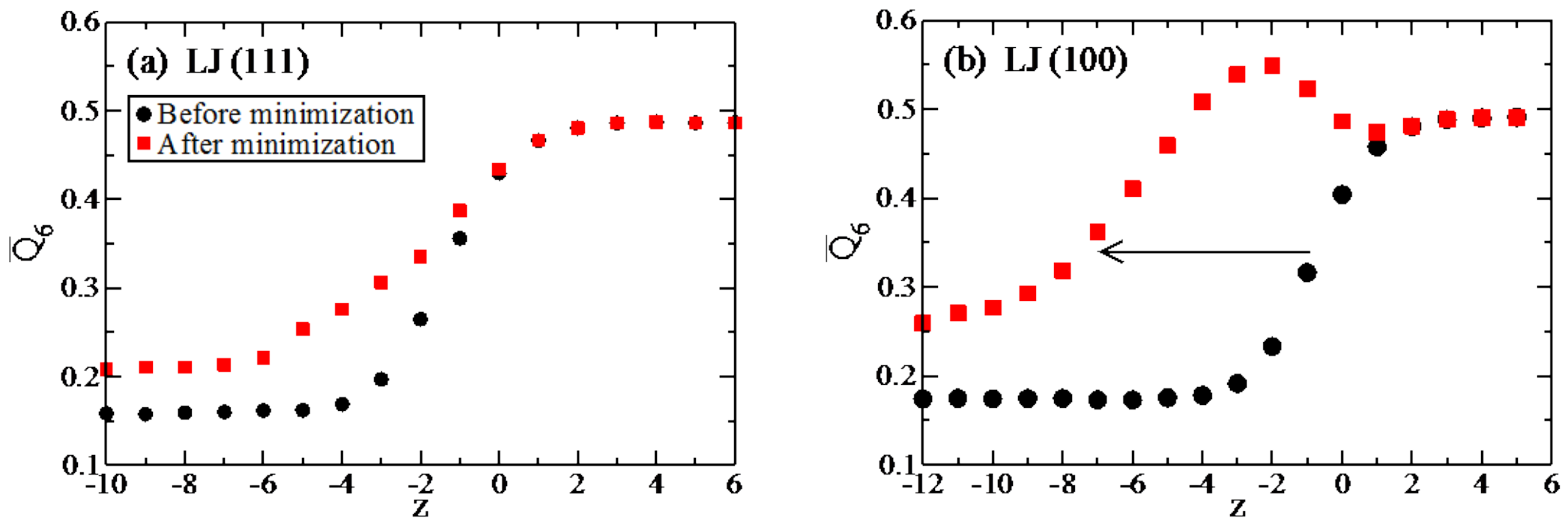

**Figure S2.** The crystal-liquid interface (111) (a) and (100) (b) of Lennard-Jones FCC crystal, as measured by $\overline{Q}_6(z)$ before and after potential energy minimization for T=0.58, averaged over by 20 interfaces. For the interface (111), there is no new crystal formed during energy minimization. While, for the interface (100), the advance of the interface corresponds to about 5 crystal layers.

The analogous representation of the change in the Cu (111) interface on minimization is shown in Fig. S3.

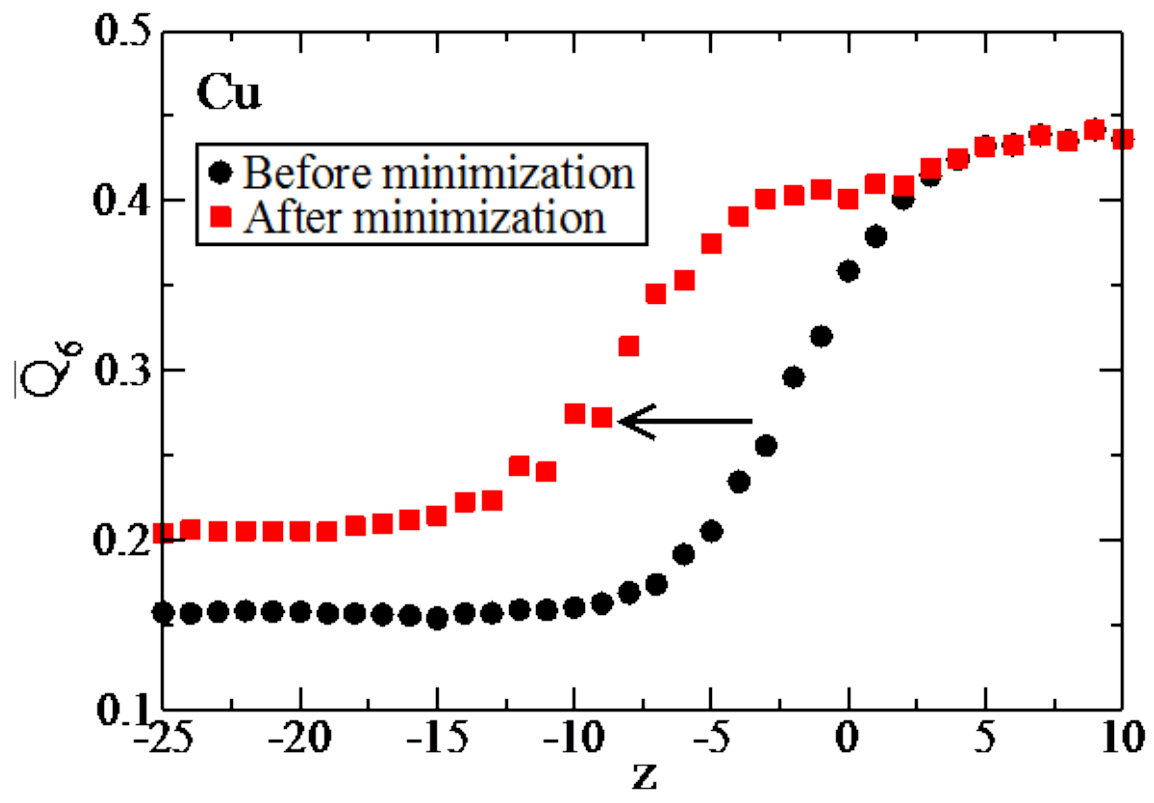

**Figure S3.** The Cu crystal-liquid interface as measured by $\overline{Q}_6(z)$ before and after potential energy minimization for T=1260K, averaged over by 20 interfaces. The advance of the interface corresponds to 3 crystal layers.

As shown in Fig. S4. the extent of the crystalline portion of the interfacial inherent structure in Cu, does not change significantly as we increase the supercooling.

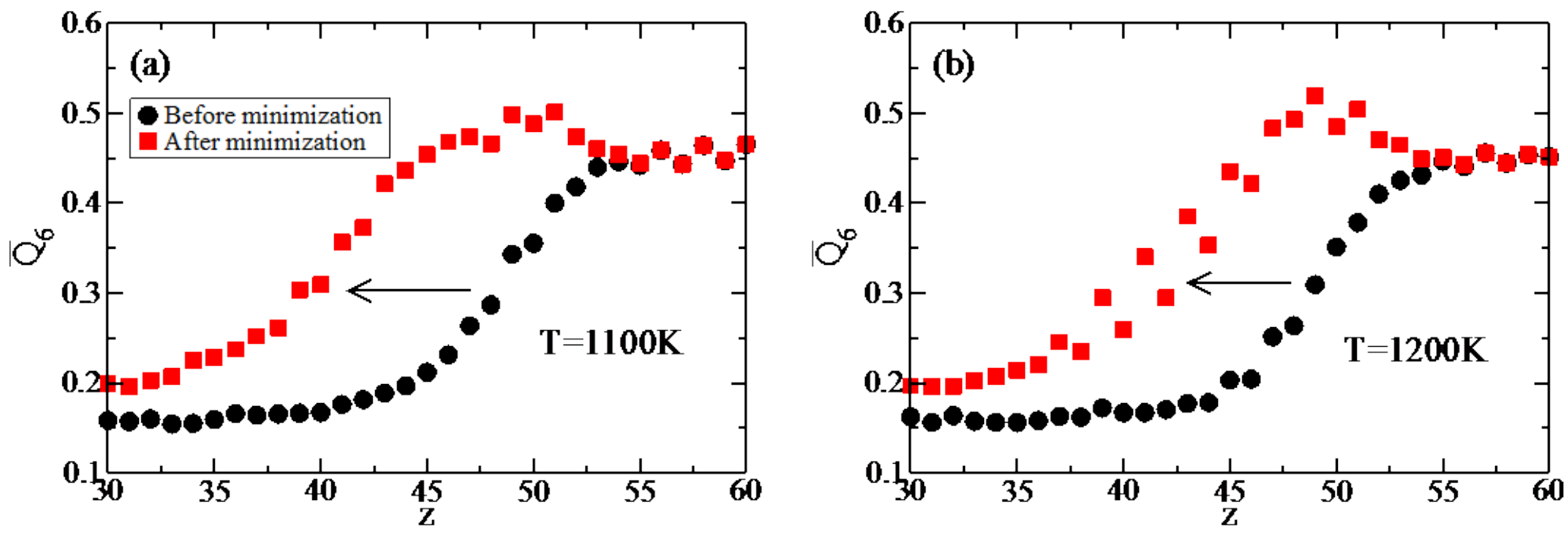

**Figure S4.** The Cu crystal-liquid interface as measured by $\overline{Q}_6(z)$ before and after potential energy minimization for (a) T=1100K and (b) 1200K. The advance of the interface corresponds to 3 crystal layers.

In Fig. S5 we have plotted the change in the interface on minimization for the other 5 metals studied in this paper, finding similar behavior in all cases.

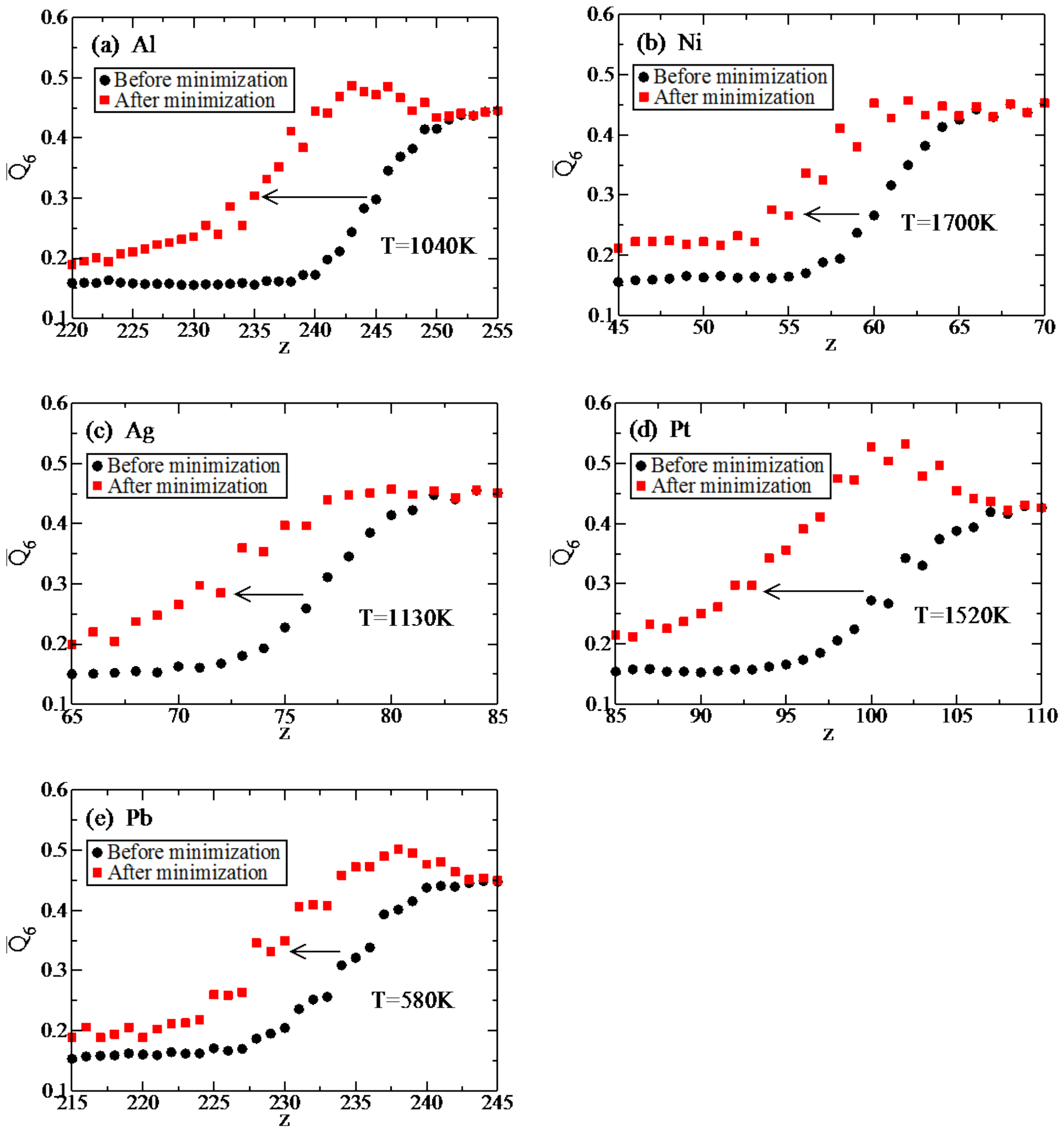

**Figure S5.** The crystal-liquid interfaces as measured by $\overline{Q}_6(z)$ before and after potential energy minimization for metals Al, Ni, Ag, Pt and Pb at temperatures close to the melting.

In Fig. S6, we show the particle displacements, analogous to that shown for Cu in Fig. 4, for the other 5 metals. Again, the behavior is similar to that found in Cu.

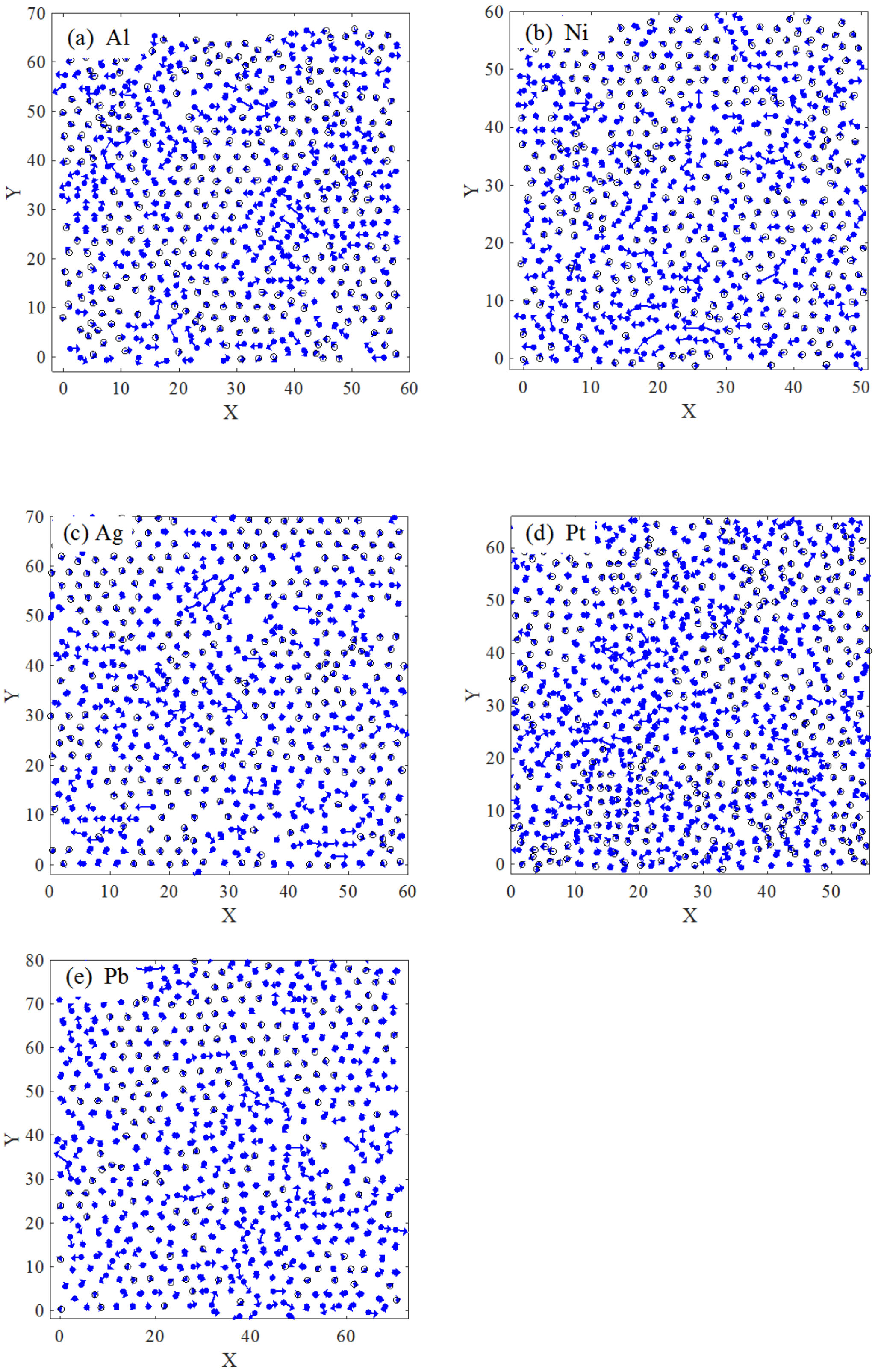

**Figure S6.** A representation of particle positions in a layer of the interface and the displacements associated with ordering during energy minimization for metals Al, Ni, Ag, Pt and Pb.

**5. Additional Data on the Turnover in the Crystal Growth Rate and the Instability of Liquid.**

In Fig. S7, we show the average crystal growth rates vs temperature of the surfaces (111), analogous to that shown for Cu in Fig. 6a, for the other 5 metals. It is found that similar Cu, the turnover of crystal growth rates, $T_{max}$, are well consistent with the kinetic spinodal temperature $T_{sp}$ .

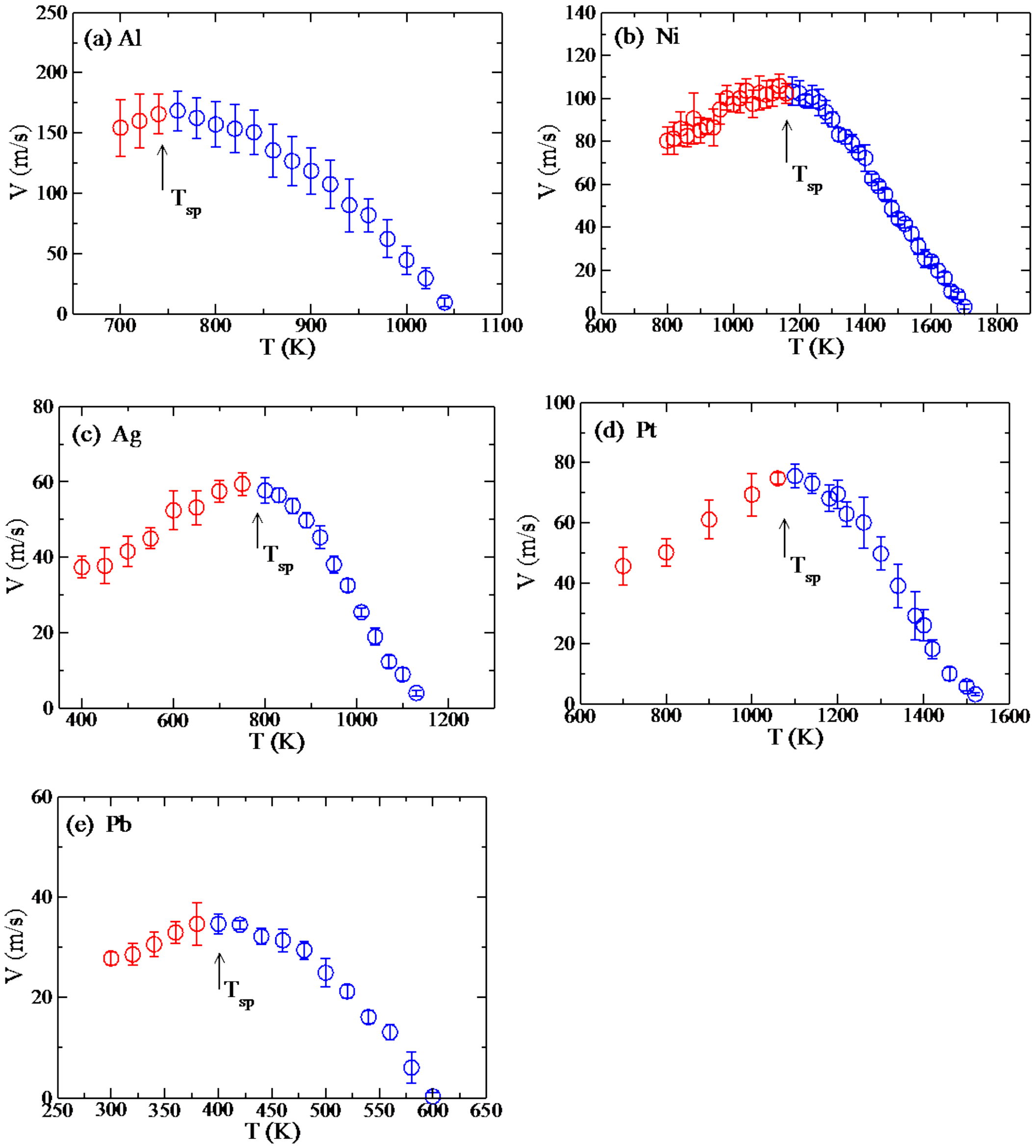

**Figure S7.** Plots of the average crystal growth rates vs temperature of the surfaces (111) for Al (a), Ni (b), Ag (c), Pt (d) and Pb (e). The crystal growth rates above the $T_{sp}$, are represented by blue circles, and below $T_{sp}$ are marked by red circles.

**6. Additional Data on the Comparison between Our Crystal Vibration Prediction and the Other Theories.**

**Table S3** The structure factors S($K_i$), relaxation time $\tau_L(K_i)$, and the correlation length $\xi_b$ of liquid at melting temperatures for all six metals.

| | **Al** | **Ni** | **Cu** | **Ag** | **Pt** | **Pb** |
|---|---|---|---|---|---|---|
| S($K_i$) | 2.86 | 2.96 | 2.31 | 2.3 | 3.07 | 2.8 |
| $\tau_L(K_i)$(ps) | 0.23 | 0.24 | 0.34 | 0.48 | 0.54 | 1.26 |
| $\xi_b$ ($\times 10^{-10}$ m) | 2.3 | 2.68 | 1.7 | 2.76 | 3.52 | 5.2 |

In Fig. S8, we show the temperature dependence of structure factors S($K_i$), relaxation time $\tau_L(K_i)$ for all six metals.

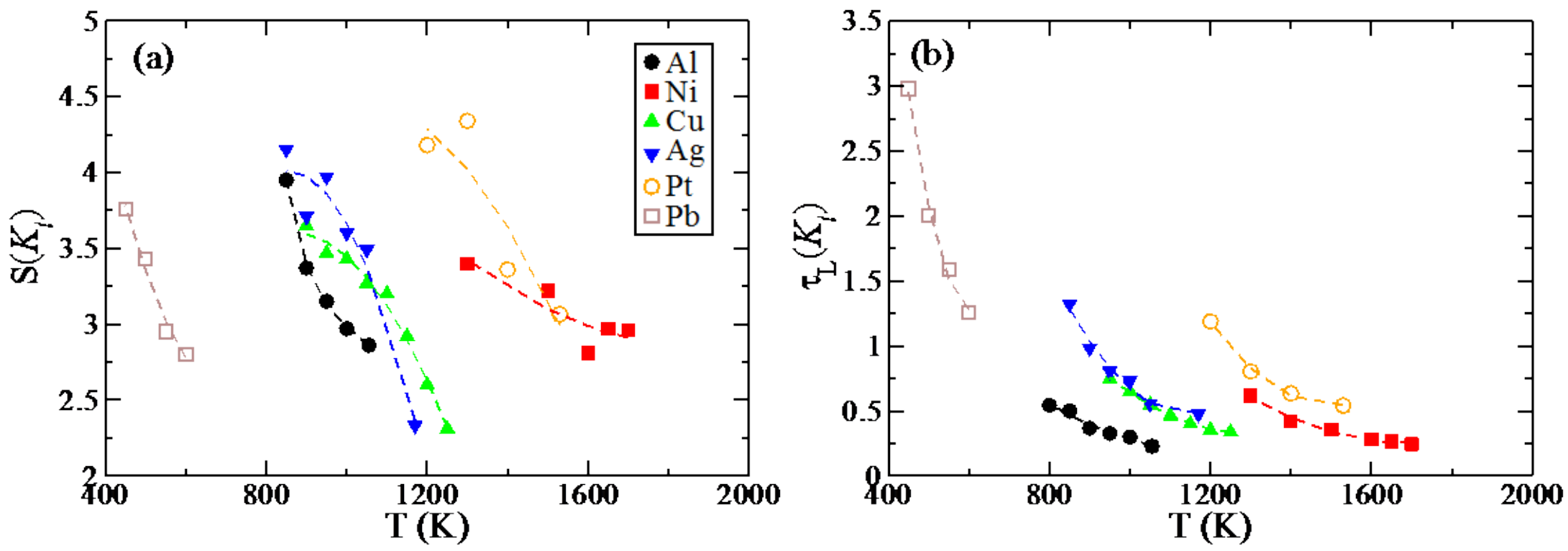

**Figure S8.** Plot of the temperature dependence of structure factors S($K_i$) (a), relaxation time $\tau_L(K_i)$ (b) for all six metals.

**7. Additional Data on the Liquid Structure before the crystal-liquid interfaces.**

In Fig. S9, we show the number density profile and orientational order profile of liquid in front of crystal-liquid interface for $T<T_{sp}$. Fig.S9(b) shows the spatial distribution of atoms with ordered configurations. These two figures supported that the nucleation occurs in liquid in front of crystal-liquid interface and the orientation of grains are random rather than parallel to the crystal surface.

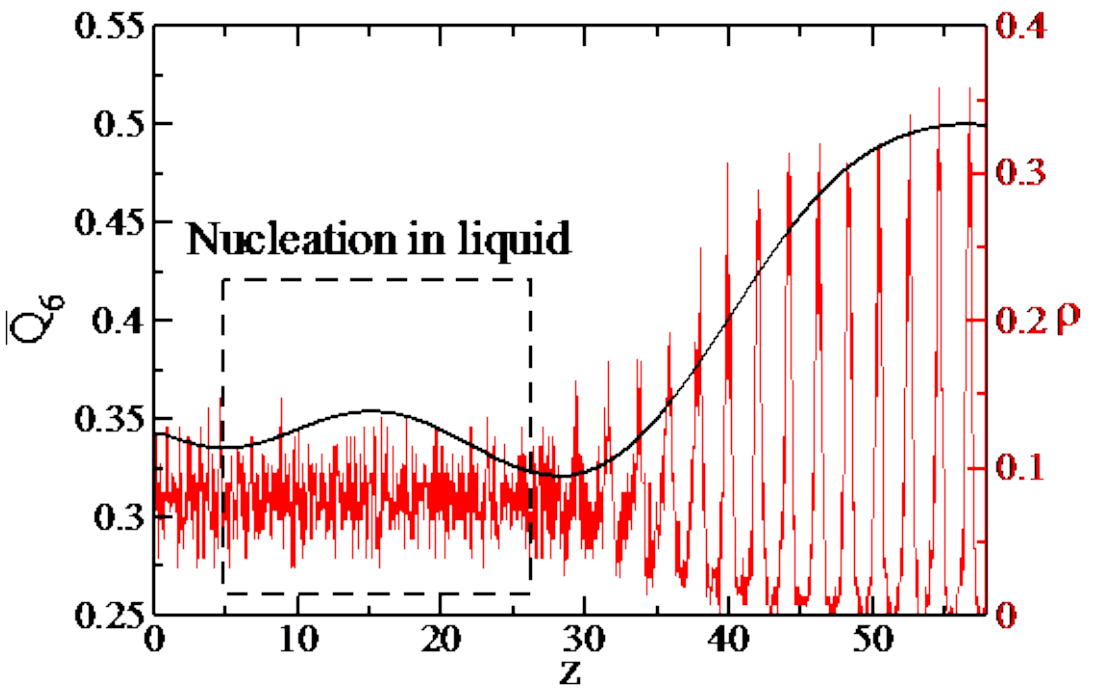

**Figure S9.** Plot of the number density profile and orientational order profile before crystal-liquid interface at for T=700K for metal Cu.

Fig. S10 shows that the variations in $Q_6$ in the grown crystal are due to (i) the particle vibration (ii) the defects formed during fast crystal growth. The defect is hcp-like crystal, according to the $Q_6$ vs $Q_4$ map.

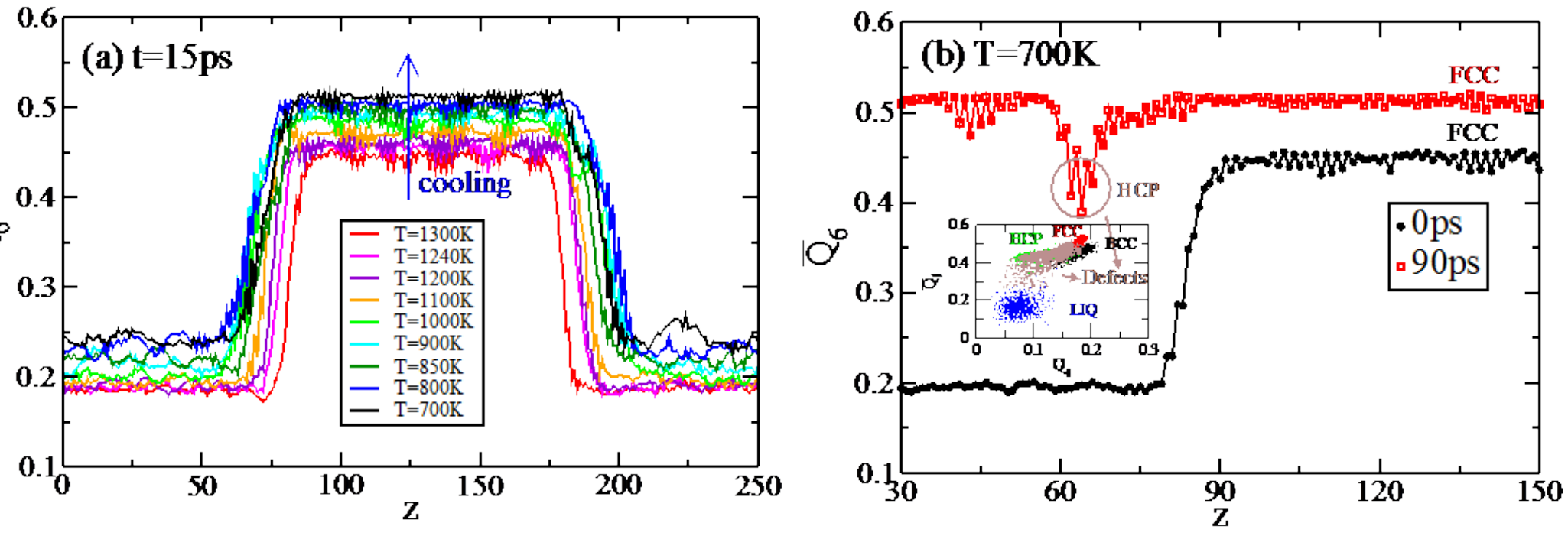

**Figure S10**. The orientational order profile (a) at the same time for different temperatures and (b) at different time for the same temperature for metal Cu. Insert: the distribution of defects' orentational order parameters in $Q_6$ vs $Q_4$ map.

In Fig. S11, we show that the crystal-liquid interface for Cu (111) includes both BCC-like and HCP like ordering, according to the distribution of the orientational order parameters of interface particles in $Q_6$ vs $Q_4$ map.

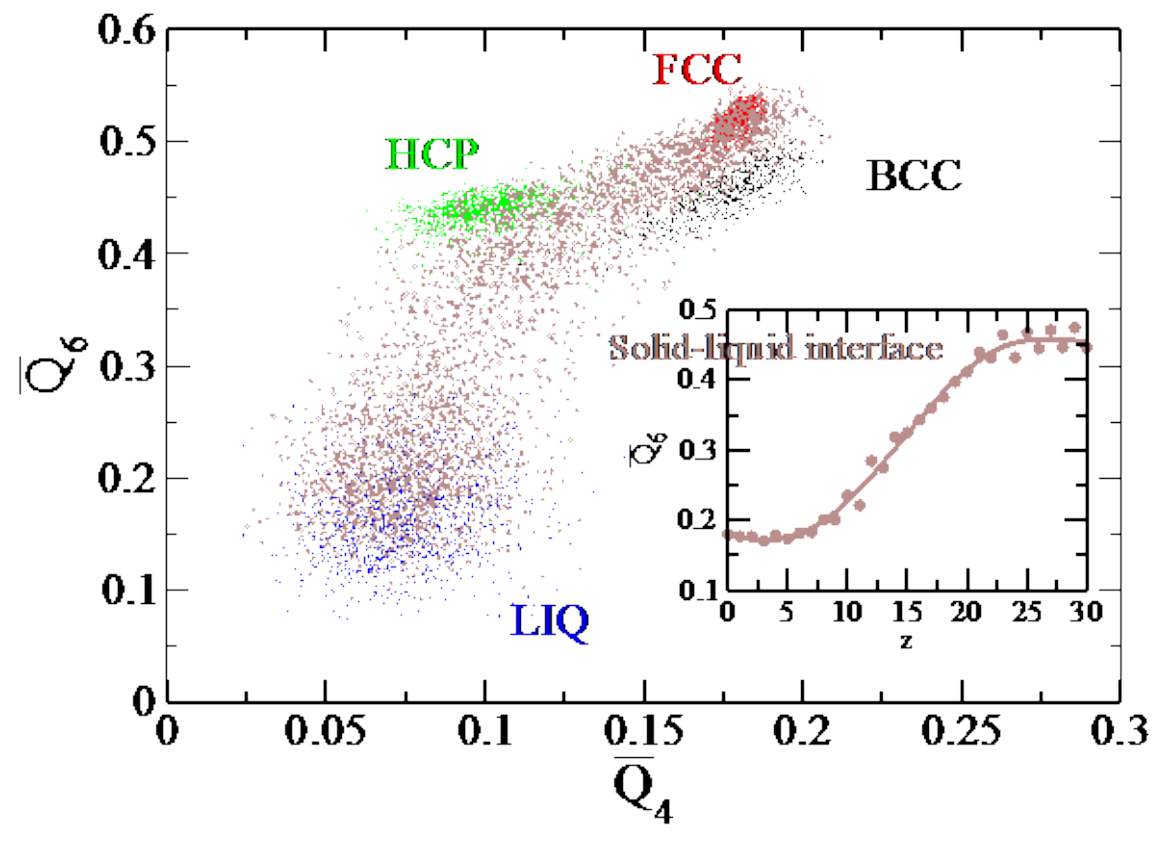

**Figure S11**. The distribution of orientational order parameters of the crystal-liquid interface particles in $Q_6$ vs $Q_4$ map for metal Cu. Insert: the $Q_6$ profile of the solid-liquid interface for Cu at T=1000K.

In Fig. S12, we show the dependence of diffusion constant and the fraction of atoms have icosahedral configurations on temperature for Cu.

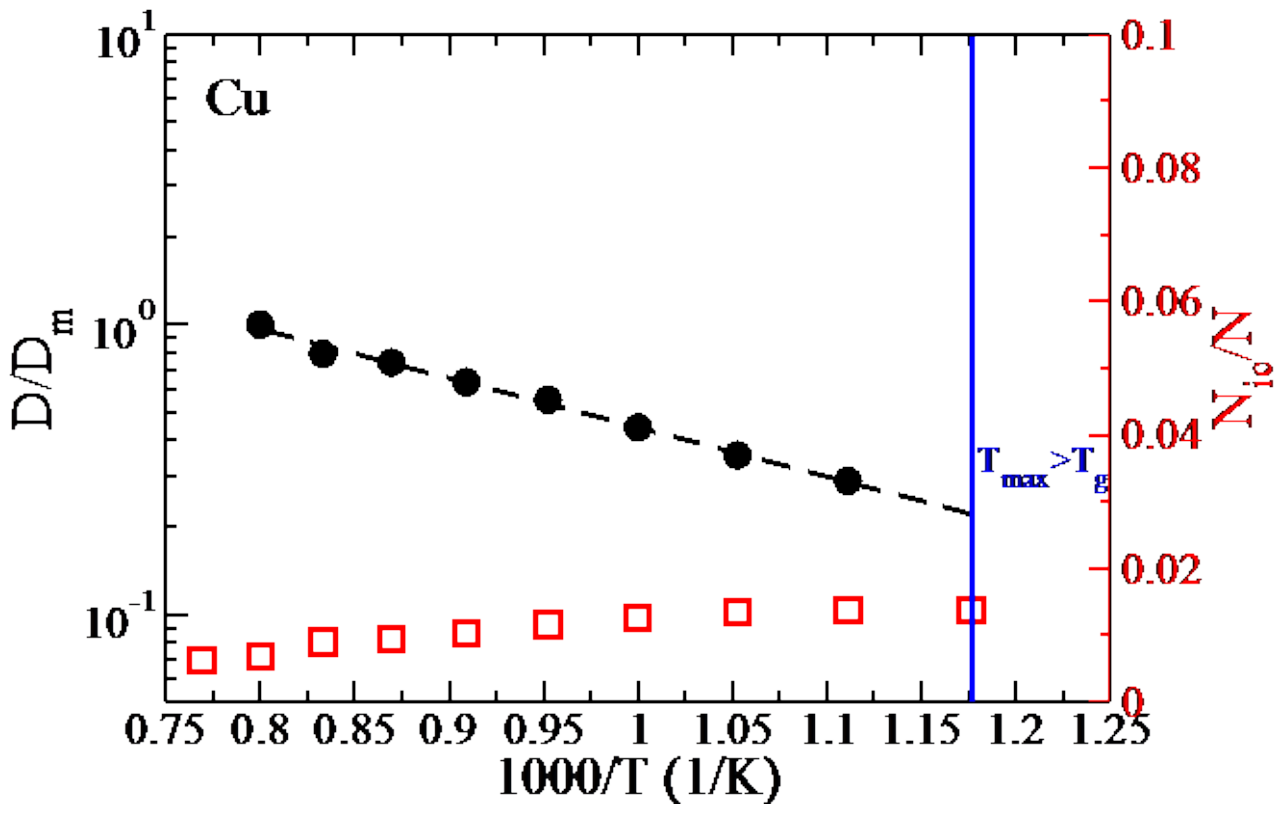

**Figure S12**. Dependence of liquid diffusion coefficients relative to the value at $T_m$ and the fraction of atoms have icosahedral configurations on temperature for Cu. The blue solid line corresponds to the turnover temperature of the crystal growth velocity for Cu. The black dashed line is the fitting curve by function y=6.22*exp(-3.9*x). For Cu, the glass transition temperature Tg is about 470~350K, much smaller than the $T_{max}$. We can see that above $T_{max}$, the dynamics follows Arrhenius function with a constant activation energy, and there is no icosahedral configurations increase sharply.